\documentclass[aps,prb,twocolumn,groupedaddress,floatfix,eqsecnum,showpacs]{revtex4}

\newcommand{\state}[1]{\ensuremath{\left|#1\right\rangle}}
\newcommand{\astate}[1]{\ensuremath{\left\langle #1 \right|}}

\newcommand{\exptvalue}[1]{\ensuremath{\left\langle #1 \right\rangle}}

\newcommand{\abs}[1]{\ensuremath{\left\vert #1 \right\vert}}

\newcommand{\Z}{\ensuremath{\textrm{\bf Z} } }

\usepackage{amsmath,amssymb,amsfonts,slashed,graphicx,hyperref}

\begin{document}

\title{Quantum noise and entanglement generated by a local quantum quench}
\author{Benjamin Hsu}
\affiliation{Department of Physics, University of Illinois,
1110 W. Green St., Urbana IL 61801-3080, U.S.A.}

\author{Eytan Grosfeld}
\affiliation{Department of Physics, University of Illinois,
1110 W. Green St., Urbana IL 61801-3080, U.S.A.}

\author{Eduardo Fradkin}
\affiliation{Department of Physics, University of Illinois,
1110 W. Green St., Urbana IL 61801-3080, U.S.A.}

\date{\today}

\begin{abstract}
We examine the growth of entanglement under a  quantum quench at point contacts of simple fractional quantum Hall fluids and its relation with the measurement of local observables. Recently Klich and Levitov proposed that the noise generated from a local quantum quench provides a measure of the entanglement entropy. Their methods were specific to non-interacting electrons and the generalization to interacting systems was left as an open question. In this work, we generalize their result to the Laughlin states. We investigate the noise generated in the current along the edge of a fractional quantum Hall state at filling factors $\nu=1/m$, when a quantum point contact, initially closed, is fully opened at some initial time $t_0=0$. We find that local quenching in these systems gives time dependent correlation functions that have universal behavior on sufficiently long time and length scales. We calculate the noise and full counting statistics for $\nu=1/m$ and find that in general, the entanglement entropy and noise generated are unrelated quantities. We also discuss a generalization of this problem to the critical quantum Ising spin chain.
	
\end{abstract}

\maketitle

\section{Introduction}

Quantum impurity problems in 1+1 dimensions have long been the subject of intense study due to numerous applications in fields such as quantum wires, the quantum Hall effect and the Kondo problem. In particular, much research has been invested into understanding the behavior of particle transfer through an impurity and of growing interest experimentally and theoretically are the statistics of the fluctuations in the transferred charge. Most efforts have been concentrated on the shot noise which is sensitive to the quantization of charge and can give useful information about the system.\cite{lesovik1989, kane1994,reznikov1995, picciotto1997, saminadayar1997,grosfeld2009} In recent years, it was realized that the full counting statistics, probability distribution $P(q)$ of the transmitted charge $q$ in a given time window $\Delta t$, contains more information.\cite{levitov1993} Along with the shot noise, the full counting statistics also contains information about the higher order correlations which are of interest experimentally. It has been suggested that the third moment may give a more reliable measure of the charge than the shot noise.\cite{levitov2004} Recent experimental efforts suggest that the higher moments of the full counting statistics may be accessible.\cite{zhang2009, gershon2008,gustavsson2006,reulet2003}

In cases where Fermi liquid theory applies, the full counting statistics has been studied extensively (see Ref. [\onlinecite{levitov_article}] for a review). Systems of strongly interacting electrons, though of importance in 1+1 dimensional quantum wires, the quantum Hall effect and Kondo problem have been less well studied. Using advanced methods such as the thermodynamic Bethe ansatz, for the Laughlin states exact results for the charge current through an impurity,\cite{FendleyLudwigSaleur-conductance} as well as the noise,\cite{FendleyLudwigSaleur-noise} the heat current, \cite{KaneFisher-thermal} and even attempts at the full counting statistics were reported.\cite{KomnikSaleur-full} Exploiting the power of boundary conformal field theory, the full counting statistics in double quantum dots was studied.\cite{sela2009}

A subject of recent interest is the  behavior of quantum entanglement, and in particular of the entanglement entropy, in condensed matter systems (for a recent review see Ref.[\onlinecite{Amico2008}].) The entanglement entropy has been shown to exhibit universal scaling behavior near quantum critical points.\cite{Calabrese2004,Fradkin2006,hsu-2009,Metlitski-2009,nishioka-2009}  Universal scaling behavior of the entanglement entropy  is also present in topological phases of matter, such as fractional quantum Hall fluids and their generalizations, as well as in the related topological quantum field theories.\cite{Kitaev2006,Levin2006,fradkin-2009} However, the entanglement entropy of a macroscopic quantum system is a highly non-local quantity which is difficult to measure. It has remained a challenge to find an experimentally viable protocol to measure the entanglement entropy.

Several recent results have suggested that the behavior of point contacts in quantum critical systems, and topological phases, may offer a way to measure the entanglement entropy.  Fendley et al have shown\cite{fendley2007} that the change in the entanglement entropy of topological FQH fluids at a point of constriction is related (in fact the same) to the change of the Affleck-Ludwig entropy\cite{Affleck1991} of the coupled edge states of the FQH fluid at the point contact. However, the Affleck-Ludwig entropy itself is difficult to measure. 

More recently, Klich and Levitov\cite{KlichLevitov} have shown that, at least for a system of free fermions, it may be possible to measure the growth (in time) of the entanglement entropy upon a quantum quench by monitoring the noise in the charge current through the contact. Specifically, they proposed that the second cumulant of the full counting statistics is related to the entanglement entropy. They considered the following protocol: first the two subsystems described by non-interacting fermions lying on either side of an infinite strength impurity are completely decoupled. At time $t_0=0$, the impurity is removed and the subsystems are allowed to exchange particles. Finally at time $t_1$, the impurity is reinstated and blocks the flow of particles. They find that the shot noise generated is given by
\begin{equation}
	S_{noise} = \frac{1}{3} \log\frac{\Delta t}{\tau}
	\label{eqn:levitov}
\end{equation}
where $\tau$ is some short time cutoff. 

Such a protocol is a suggestive procedure for measuring the quantum entanglement between two halves of a system. In the initial state the two subsystems lying on either side of the impurity are completely independent, but following the quench, quantum entanglement is dynamically built up by the exchange of particles with vanishing net flow. A measure of the entanglement between the two subsystems is provided by the entanglement entropy. The entanglement entropy for such a scenario was recently calculated using conformal field theory (CFT) and shown to increase logarithmically with time, with a universal coefficient proportional to the central charge $c$ characterizing the CFT, \cite{CalabreseCardy-CFT, Eisler2008,Eisler2007} 
\begin{equation}
	S_{ent} = \frac{c}{3} \log\frac{\Delta t}{a}
	\label{eqn:cardy}
\end{equation}
where $a$ is a short distance cutoff. By comparing Eqns. \ref{eqn:levitov}-\ref{eqn:cardy} and noting that for non-interacting fermions, the central charge is $c=1$, it is quite suggestive that the noise is a measurement of the entanglement. However, it should be noted that while the methods of Calabrese and Cardy are general, the results of Klich and Levitov are specific to non-interacting electrons.
The degree of general validity ({\it e.g.\/} for interacting systems)  of the Klich-Levitov protocol is presently an open question. 

Recently  there has also been a surge of interest in quantum quenching, {\it i.e.\/} where the parameters that describe the dynamics of the system are changed over a short period of time either locally or globally. In cold atoms, it has been seen experimentally that quantum quenches may show interesting features.\cite{greiner2002, altman2002, kinoshita2006} Theoretically, quantum quenching has also attracted a good deal of attention in recent years.\cite{barouch1970,sengupta,igloi2000,calabrese2005,CalabreseCardy-CFT} It was found recently that in a global quench, where the eigenstate state $\state{\psi_{0}}$ of a Hamiltonian $H_{0}$ is evolved by a different Hamiltonian $H$, the correlation functions were found to display universal behavior characteristic of the Hamiltonian $H$ if it was tuned near criticality.\cite{calabrese2006}

\begin{figure}
 \includegraphics[width=0.35\textwidth]{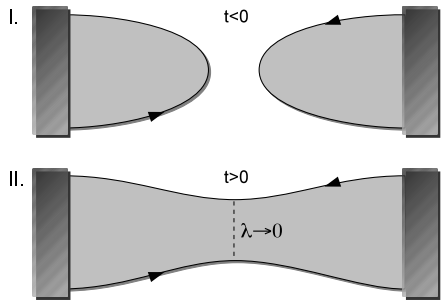}
\caption{Single quench at $t=0$. (I) Fully reflecting for $t<0$ ($\lambda\to\infty$), (II) Fully transmitting for $t>0$ ($\lambda\to 0$).}
\label{Fig:quench}
\end{figure}

In this work, we study a {\it local} quantum quench where the point contact between two $\nu=1/m$ fractional quantum Hall states is instantaneously opened (see Fig. \ref{Fig:quench}). We examine the noise and the full counting statistics and extend the results of Klich and Levitov[\onlinecite{KlichLevitov}] to a system of interacting electrons with a dynamic impurity. No external bias is applied. Here the bare strength of the impurity coupling undergoes a sudden change between two values, the first corresponding to fully reflecting boundary conditions and the second to fully transmitting boundary conditions. The quantum point contact (QPC) generates an effective impurity in the Luttinger liquid description of the edge, thus allowing complete control over the impurity strength by tuning the voltage applied to a side gate. Such a system is therefore a promising candidate for an experimental exploration of quantum quenches.

Our main result is that  the second moment of the full counting statistics, and therefore the noise, has a similar form to the entanglement entropy. This result holds even for interacting (Luttinger) systems. However, we also find that this correspondence appears to be coincidental since, in general, in addition to the noted dependence of the central charge,  the noise depends also on other universal quantities of the underlying conformal field theory of the system. There is however a conceptual difference between static and dynamic entanglement. Static entanglement refers to the measure of the non-local correlations associated with the {\em observation} of a part of the system without disturbing it and is encoded in the von Neumann entanglement entropy. On the other hand, when a physical system is changed, as in a quantum quench, the ensuing {\em dynamical }entanglement measures the non-local correlations that develop upon its time evolution in the quenched system. (See, {\it e.g.} Ref. [\onlinecite{calabrese-2009b}]). At the computational level, the main method underlying our results is the use of a boundary condition changing operator to generate the transition between the fully closed QPC and the fully open one. This transition is related to a change in the sign of the odd-boson density operator going through the QPC, thus is marked by the appearance of an orbifold theory.

The method in its most general form will be described in section \ref{section:Orbifolds}. In sections \ref{section:Majorana} and \ref{section:Orbifolds} we describe the method for $\nu=1/2$ and $1/3$. For the particular case of $\nu=1/2$, the result can be derived explicitly using a Majorana fermion representation for the re-fermionized Hamiltonian. Finally, in section \ref{section:Noise} we calculate the noise, and in \ref{section:FCS} the full counting statistics. In Section \ref{section:Thermal} we sketch the calculation of the noise of the energy-momentum in the quantum Ising chain, showing that in contrast to the entanglement entropy in a 1+1 dimensional critical system, the noise does not always grow logarithmically. The details of the calculations for the Laughlin $\nu=1/2$ (bosonic) state are presented in Appendix \ref{appendix:Nu-half}. The modular $S$-matrix and fusion rules for the orbifold CFT (used in Section \ref{section:Orbifolds}) are given in Appendix \ref{appendix:OrbifoldAlgebra}. The relation between the Schwinger-Keldysh formalism and entanglement entropy is summarized in Appendix \ref{Appendix:KeldyshEntangle}.

\section{Description of the Model}

The Lagrangian describing the edge of the quantum Hall effect for the Laughlin states at filling factors $\nu=1/m$ can be written in the terms of left and right moving bosonic fields, $\phi^{L/R}$, in the form
\begin{equation}
	\mathcal{L} = \frac{1}{4\pi} \partial_{x} \phi^{L} (\partial_t-\partial_x) \phi^{L} - \frac{1}{4\pi} \partial_{x}\phi^{R}(\partial_t+\partial_x) \phi^{R}
\end{equation}
where $x\in[-\ell,\ell]$, and the velocity $v$ of the edge was set to one. For decoupled edges, the physical Hilbert space of each chiral boson is generated by $j$, the $U(1)$ current and the vertex operator $e^{i \sqrt{g}\phi_{L/R}}$ (see Ref.[\onlinecite{wen1990,wen1995}]). This vertex operator generates the charged excited states of the theory which are fundamentally generated by adding or subtracting an electron from either the left or right edge.\cite{kane1992,wen1990} Requiring that the charged excitations generated by $e^{i\sqrt{g}\phi_{L/R} }$ are bosons (electrons) implies that they satisfy the (anti-) commutation relation sets $g=m$ an even (odd) integer. The requirement that the charge of the operator is precisely $e$ forces the relation $\nu = 1/m$. Representations of the theory are local with respect to the $U(1)$ current and charged excitation $\Psi_{L/R} = e^{i \sqrt{g}\phi_{L/R}}$ which is identified with the electron ($g=m$ odd) or boson ($g=m$ even). 

The QPC introduces backscattering at $x=0$, with the most relevant term being the quasi-particle tunneling term
\begin{equation}
	\mathcal{L}' = \lambda \delta(x) \cos \left[\sqrt{\nu}(\phi^{R}-\phi^{L})\right]
\end{equation}
where $\lambda$ is the tunneling strength. The charge densities at the left and right edges are respectively
\begin{equation}
	\begin{array}{ccc}
		\rho_{L} = \frac{\sqrt{\nu}}{2\pi } \partial_{x} \phi^{L} &\,;\,& \rho_{R} =- \frac{\sqrt{\nu}}{2\pi }  \partial_{x} \phi^{R}
	\end{array}
\end{equation}
Note that $\phi_L(x+2\ell)=\phi_L(x)+2\pi \sqrt{\nu} Q^L$, so in particular the electron operator is well defined provided $Q^L$ is an integer; hence $Q_L$ carries the meaning of the number of quasi-particles on the edge. In units of the electron charge $e$, the charge carried by the edge is $\nu Q_L$.
	
We now perform the following transformation into the odd and even basis
\begin{eqnarray}
	&& \nonumber  \phi_o(x,t)=\frac{1}{\sqrt{2}}\left(\phi_L(x,t)-\phi_R(-x,t)\right)\\
	&&\\
	&& \nonumber	\phi_e(x,t)=\frac{1}{\sqrt{2}}\left(\phi_L(x,t)+\phi_R(-x,t)\right)
\end{eqnarray}
the even boson $\phi_e$ decouples from the impurity, and we ignore it in the rest of the paper. Note that $\phi_{o/e}$ are both left-moving. The odd and even charges are related to the original $L/R$-charges through $\sqrt{2}Q^o=(Q_L-Q_R)$, $\sqrt{2}Q^e=(Q_L+Q_R)$.

The experimental setting we wish to consider is one where the two edges of the quantum Hall liquid are separated initially. No bias is applied 
and the temperature is taken to be zero. 
At some time, the two halves of the system are (suddenly) connected and we wish to consider the statistics of the charge transferred from one side to the other upon this change, {\it i.e.} a quantum quench. The system is then evolved with the Hamiltonian with a Hilbert space described by states with transmitting boundary conditions at the quantum point contact.

Let $P(q)$ be the probability that charge $q$ is transmitted through the point contact in time $\Delta t$. Then one defines the generating function $\chi(\lambda) = \sum_{q=-\infty}^{\infty} P(q) e^{i\lambda q}$ which encodes all moments (or cumulants) $C_{m}$ of the probability distribution $P(q)$,
\begin{equation}
	C_{m} = (- i \partial_{\lambda})^{m} \log \chi(\lambda) \bigg\vert_{\lambda=0}
\end{equation}
In particular, the second moment $C_{2}$ is related to the current fluctuations. The generating function can be written as
\begin{equation}
	\chi(\lambda)=\left\langle\left\{e^{i \lambda \hat{q}(\Delta t)},e^{-i\lambda \hat{q}(0)}\right\}\right\rangle
\end{equation}
where the operators are ordered on the Schwinger-Keldysh contour.\cite{levitov1993} This can also be written as a trace over a complete set of states. 
\begin{equation}
	\chi(\lambda) = \textrm{Tr} \left(\rho_{0} U^{\dagger} e^{i\lambda q} U e^{-i\lambda q} \right)
	\label{eq:chi-lambda}
\end{equation}
where $\rho_{0}$ is the {\it initial} density matrix, and $U$ is the time evolution operator for time $\Delta t$ after the point contact is opened. In Eq.\eqref{eq:chi-lambda} it should be understood that the trace is taken over the initial states. 

If the scattering time at the quantum point contact is short compared with the entire time evolution, then it was shown that $\chi(\lambda)$ can be written as\cite{levitov1993, klich_article}
\begin{equation}
	\chi(\lambda) = \textrm{det} \left( 1 + n ( S^{\dagger} e^{i\lambda q} S e^{-i\lambda q} -1) \right)
\end{equation}
where $n$ is the initial distribution of states and $S$ is the reflection-transmission matrix at the quantum point contact. 
In Ref.[\onlinecite{KlichLevitov}], it was found that for {\em non-interacting electrons}, the generating function is given by
\begin{equation}
	\chi(\lambda) = e^{-\frac{\lambda^{2} }{2}  C_{2} }
\end{equation}
where $C_{2} = \frac{1}{\pi^{2} } \log\left( \frac{\Delta t}{\tau} \right)$. 

In this work, we generalize this protocol to the $\nu=1/m$ Laughlin states. We note that while the full counting statistics has been found in many other strongly correlated systems with a static impurity via the thermodynamic Bethe ansatz,\cite{gogolin2007, KomnikSaleur-full, FendleyLudwigSaleur-conductance,FendleyLudwigSaleur-noise} the application of this method to finding time and space dependent correlation functions is in general difficult and remains largely not understood.

We take a different approach in tackling this problem. This relies on the realization that one should consider an extended Hilbert space by introducing a boundary changing operator to the edge theory and that physical charged excitations in the initial Hilbert space are those conserving total charge. In the next section, we examine the noise in $\nu=1/2$ which can be refermionized and solved explicitly. We note an interesting structure which we then extend to the other Laughlin states at $\nu=1/m$. In doing so, we calculate the noise and full counting statistics for the Laughlin states.

\section{Noise at filling factor 1/2}

\label{section:Majorana}

We start our discussion with an exactly solvable case, the bosonic quantum Hall effect at filling factor $\nu=1/2$. Our strategy is as follows: re-fermionizing, we first recast the problem in terms of a quadratic fermionic action. Then, by writing the fermionic field in terms of two real Majorana fermion fields, we find that only one of the Majorana fermions interacts with the impurity. The two fixed points associated with the fully transmitting and the fully reflecting impurity, translate into anti-periodic and periodic boundary conditions imposed on this Majorana fermion. This leads us to identify the operator that takes the system between the two fixed points as the $\sigma$ operator of the chiral Ising model. Armed with this knowledge, we then proceed to use this formalism to calculate the noise for the case of a quantum point contact which is suddenly opened at time $t=0$. 
 
\subsection{Majorana fermion representation}

The odd boson theory can be re-fermionized in terms of a fermion field $\psi$ (see Appendix \ref{appendix:Nu-half}) leading to the Hamiltonian
\begin{eqnarray}
	\nonumber H_o&=&\int dx\, \psi^\dag(x)i\partial_x\psi(x)\\
	&&\quad\quad +\sqrt{2\pi}\lambda\delta(x)\left[\psi(x)\gamma+ \gamma\psi^\dag(x)\right]
\end{eqnarray} 
The crucial step is that the Hamiltonian can be written in a Majorana representation, $\psi=(\eta_1-i\eta_2)/2$ where $\eta_1$ and $\eta_2$ satisfy $\{\eta_i(x),\eta_j(x')\}=2\delta_{ij}\delta(x-x')$. In this representation, only $\eta_2$ interacts with the impurity, as is clear by inspecting the equations of motion,
\begin{eqnarray}
	\nonumber &&i\partial_t i\eta_1=i\partial_x i\eta_1\\
	&&i\partial_t i\eta_2=i\partial_x i\eta_2+2\sqrt{2\pi}\lambda \gamma\delta(x) \label{eq:mot2}\\
	\nonumber &&i\partial_t\gamma=2\sqrt{2\pi}\lambda i\eta_2(0) 
\end{eqnarray}
Integrating the equations leads to the following boundary condition across the impurity, $\omega_B\equiv 4\pi \lambda^2$,
\begin{eqnarray}
	\eta_2(0^+)=\frac{i\omega+\omega_B}{i\omega-\omega_B}\eta_2(0^-)
\end{eqnarray}
This boundary condition is continuously interpolating between the fully reflecting and fully transmitting boundary conditions for the odd boson
\begin{eqnarray}
	\begin{array}{lll}
	\omega_B\to\infty & \quad\eta_2(0^+)=-\eta_2(0^-)\\
	\omega_B\to 0 & \quad \eta_2(0^+)=\eta_2(0^-)
	\end{array}
\end{eqnarray}
Similarly, at $x=\ell$, we have the boundary condition,
\begin{eqnarray}
	\eta_2(\ell)=\eta_2(-\ell)
\end{eqnarray}
The problem is thus mapped to a single chiral Majorana fermion on a circle of circumference $2\ell$. For the reflecting case we have anti-periodic (AP) boundary conditions, while for the transmitting case, we have periodic (P) boundary conditions.

The free Majorana field theory contains altogether three primary fields: the identity field {\bf 1}, the Majorana fermion field $\eta$, and a $\sigma$ operator (known as the Ising twist field) creating a branch cut on the Majorana fermion $\eta$. As before, we shall label the two copies of fields present here by an index $i=1,2$: {\bf 1}$_i$, $\eta_i$, and $\sigma_i$. To go from periodic to anti-periodic boundary conditions on $\eta_2$, we use the $\sigma_2$ operator to create a branch cut on the fermion: we imagine that the edge Majorana fermion encloses a ``bulk" in which $\sigma_2$ operators can be introduced, appropriately changing the boundary conditions on the Majorana fermion. We thus identify the operator taking the system between the two fixed points associated with the QPC as the $\sigma_2$ operator in the $\eta_2$ Majorana fermion field theory.

\begin{widetext}

\begin{figure}
  \includegraphics[width=0.8\textwidth]{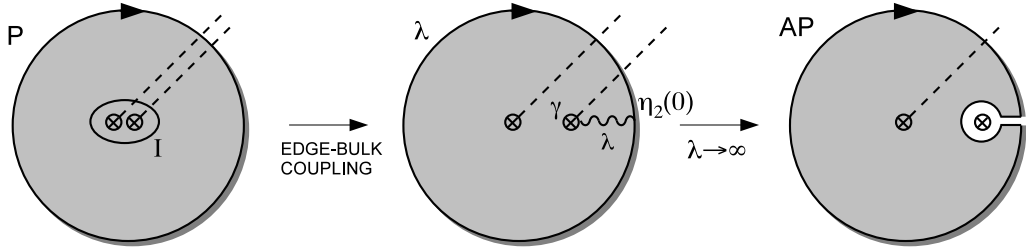}
\caption{Boundary changing operator - fully transmitting to fully reflecting. Two $\sigma_2$ operators are drawn from the vacuum. Tunneling (of strength $\lambda$) is introduced between one of the $\sigma_2$ operators and the edge. Finally, in the limit $\lambda\to\infty$, the edge circumvents the $\sigma_2$ operator.}
\label{fig:boundary-changing}
\end{figure} 

\end{widetext}

In particular, to go from anti-periodic to periodic boundary condition, introduce two $\sigma_2$ operators in the ``bulk" surrounded by the $\eta_2$ edge (see Fig. \ref{fig:boundary-changing}). One of them then approaches the edge at $x=0$ and a bulk-edge coupling is introduced. This coupling is mediated by virtual processes involving the tunneling of topologically trivial modes between the bulk and the edge \cite{grosfeld2006}. Finally, when $\lambda\to\infty$, the edge circumvents the operator completely, and switches to AP boundary conditions. In the sudden approximation, we can imagine two spatially-separated $\sigma_2$ operators introduced simultaneously: one inside the edge and the other outside.

\subsection{Calculation of the noise}

We assume that the QPC is initially closed, then at time $t=0$ it is fully opened, and proceed to calculate the noise produced in that situation. 

The current operator through the junction can be defined as the rate of transfer of charge between the two edges, measured by the odd boson
\begin{eqnarray}
	I=\partial_t\rho_o=i[H,\rho_o]
\end{eqnarray}
where $\rho_o=\psi^\dag \psi$. Using the equations of motion, Eqns. \eqref{eq:mot2}, we can write the current in terms of the Majorana fermions as\begin{eqnarray} \label{eqn:current}
	I=\frac{i}{2}\eta_1(0)(\eta_2(0^+)-\eta_2(0^-))
\end{eqnarray}
In the limits, $\omega_B\to \infty$ and $\omega_B\to 0$, we get
\begin{eqnarray}
	I_{b}=\left\{\begin{array}{ll}
	0 & t<0\\
	i\eta_1(0)\eta_2(0^+) & t>0\\
	\end{array}\right.
\end{eqnarray}
where $t=0$ time of the quench, and $I_{b}$ is what we call the backscattered current.

In the following, we shall define the charge transmitted through the junction during a time window starting at $t=-\infty$ and ending at some arbitrary positive time $\Delta t$,
\begin{eqnarray}
	Q_{\Delta t}=\int_{-\infty}^{\Delta t}dt\, I_{b}(t)=\int_0^{\Delta t} dt \,I_{b}(t)
\end{eqnarray}
Since there is no bias, the net transferred charge is zero, $\langle Q_{\Delta t}\rangle=0$. However, there will be noise due to the opening of the QPC. The noise is given by
\begin{eqnarray}
	\langle Q_{\Delta t}^2 \rangle=\int_0^{\Delta t} dt_1\int_0^{\Delta t}dt_2 \frac{\langle\sigma_2(0)I_{b}(t_1)I_{b}(t_2)\sigma_2(0)\rangle}{\langle\sigma_2(0)\sigma_2(0)\rangle}
\end{eqnarray}
Besides its role in establishing a non-zero current operator, the operator $\sigma_2$ also changes the boundary conditions on $\eta_2$. However, these boundary conditions ({\it i.e.\/} the effects of the operator $\sigma_2$)  will affect $\eta_2$ only after a long time-scale set by $\ell/v$. In the limit $\ell\to\infty$ we can safely neglect the effects of these boundary condition changing operators. Upon the introduction of a UV cutoff $\delta$, the short-time switching scale needed to regulate the integral,  the second cumulant $C_{2}$ can be obtained.
\begin{eqnarray} \label{eqn:2ndcumulant}
	\langle Q^2_{\Delta t} \rangle=\frac{\nu}{2\pi^2}\int_0^{\Delta t} dt_1\int_0^{\Delta t}dt_2\frac{1}{(\delta+i(t_1-t_2))^2}
\end{eqnarray}
The expression can be easily integrated to get
\begin{eqnarray}
	\langle Q_{\Delta t}^2\rangle=\frac{\nu}{2\pi^2}\log\frac{\Delta t^2+\delta^2}{\delta^2}
\end{eqnarray}
In the limit of large $\Delta t$ this reduces to
\begin{eqnarray}
	C_2=\frac{\nu}{\pi^2}\log\frac{\Delta t}{\delta}
\end{eqnarray}

The essential point is that the theory of the two edge states coupled by the local QPC  must have a local boundary changing operator $\sigma_2$ that takes the system from one fixed point to the other. In particular, the operator $\sigma_2$ maps the ground state of the system with a closed QPC  to an energy eigenstate of the same system but with the new boundary conditions, an open QPC. The ensuing growth of the entanglement entropy is due to the evolution of correlations of the excitations in this state.\cite{calabrese-2009b}

\subsection{Bosonic Picture}

The previous discussion can also be understood in terms of the boson. This picture lends itself to generalization to the other Laughlin states which are also described by free chiral bosonic theories. Here, we show that from the bosonic picture, one can deduce that $\sigma$ operator must be introduced for one of the Majorana fermions to change the boundary condition from anti-periodic to periodic. We then argue that a similar boundary condition changing operator can be included in the bosonic theory.

One can understand the relationship between boson and fermion boundary conditions and the nature of the twist operator by examining the partition functions. Specializing to $\nu=1/2$, the Lagrangian becomes
\begin{equation}
	\mathcal{L} = \frac{1}{4\pi } \partial_{x}\phi_{o} ( \partial_{t}-\partial_{x} ) \phi_{o} + \lambda \delta(x) e^{i  \phi_{o}(x) } + \textrm{h.c.}
\end{equation}
The boundary condition at the impurity site changes from $\lambda\rightarrow \infty$ perfectly reflecting (Neumann boundary conditions) to $\lambda\rightarrow 0$ perfectly transmitting (Dirichlet boundary conditions). The Neumann boundary condition simply fixes the field $\phi_{o}$ to be continuous across the boundary while the Dirichlet boundary condition allows the field to be discontinuous.

Imposing Neumann boundary conditions at the ends of the system for all times, the mode expansion for the odd boson can be found. For Neumann (N) and  Dirichlet (D) boundary conditions at the origin respectively,
\begin{eqnarray}
 	\nonumber	&& \phi_{o}(x,t) = 
	\phi_{0}+  \frac{ \pi \nu}{\ell} Q^{o}(t-x)
	\nonumber \\
	&& \;\;\;\;\; \;\;\;\;\;\;\;\;\; +  i\sqrt{\frac{\nu}{2} }\sum_{n\in \Z} \frac{\alpha_{n}}{\sqrt{\vert n\vert } } e^{-\frac{i \pi}{\ell}n(t-x)}      \qquad (N) \nonumber \\
	&&  \\
	\nonumber 	&&  \phi_{o}(x,t) = i\sqrt{\frac{\nu}{2} } \sum_{r\in \textrm{\bf Z} + \frac{1}{2} } \frac{\alpha_{r}}{\sqrt{\vert r\vert} } e^{-\frac{i\pi}{\ell}r(t-x)} \qquad (D)
\end{eqnarray}
The partition function in each sector can then be computed for each set of boundary conditions and one finds
\begin{eqnarray}
	\nonumber	 \lambda\rightarrow \infty \quad &\Rightarrow &\quad Z_{NN} = \frac{\theta_{3} (q) }{\eta(q) } =Z^{f}_{AA} Z^{f}_{AA} \\
	&& \\
	\nonumber	\lambda\rightarrow 0 \quad &\Rightarrow &\quad  Z_{ND} = \sqrt{ \frac{\theta_{2} (q)\theta_{3}(q) }{\eta^{2}(q) }} = Z^{f}_{PA} Z^{f}_{AA}
\end{eqnarray}
where $Z^{f}_{ij}$ are partition functions for a single $c=\frac{1}{2}$ holomorphic fermion with boundary conditions $i$ in time and $j$ in space, A anti-periodic and P periodic. To go from $\lambda\rightarrow \infty$ to $\lambda\rightarrow 0$, we see that exactly one Majorana fermion changes anti-periodic to periodic boundary conditions in time. Now, the partition function $Z^{f}_{PA}$ is also related to the partition function $Z^{f}_{AA}$ by an insertion of the fermion counting operator $(-1)^{F}$ in the time direction.  By a conformal transformation, this operator is related to the spin operator of the Ising model inserted in the space direction for one of the Majorana fermions that compose the theory.\cite{ginsparg1990} This confirms the results of the previous section.

While this identifies the boundary changing operator as the spin operator in the Ising model, this is a special feature of $\nu=1/2$ state. At other filling fractions the boson can no longer be fermionized, but what is independent of the filling is the boson mode expansion and the partition functions $Z_{ND}, \, Z_{NN}$. To generalize the calculation of the noise beyond $\nu=1/2$, one should identify the boundary condition changing operator in the bosonic theory that takes Neumann boundary conditions to Dirichlet boundary conditions.

\begin{figure}
 \includegraphics[width=0.47\textwidth]{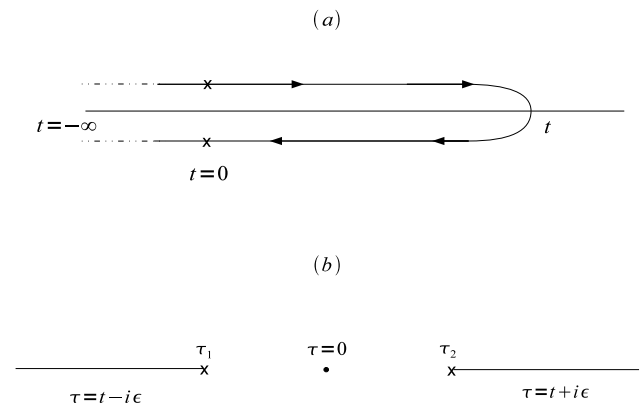}
\caption{(a) The Schwinger-Keldysh contour can be regarded as the time contour of a conformal field theory by an analytic continuation. (b) The path integral can be thought of as over the entire surface parameterized by $(x,\tau)$ with changing boundary conditions at $\tau_{1}$ and $\tau_{2}$. Details in Appendix \ref{Appendix:KeldyshEntangle} and [\onlinecite{calabrese2005,calabrese2006}].}
\label{fig:keldysh}
\end{figure}

A priori, it may seem strange that a $c=1$ free bosonic theory must include a boundary changing operator, but the necessity of this operator can be seen at the level of the path integral. The path integral should be time ordered on the Schwinger-Keldysh contour, but by an analytic continuation, one can regard the Schwinger-Keldysh contour as the time surface for a conformal field theory\cite{calabrese2005} (see Fig.\ref{fig:keldysh}). In going from the density matrix to the path integral formulation, one must consider an {\it extended} Hilbert space of the Hamiltonian $\mathcal{H}$ which is spanned by states with all possible boundary conditions. The boundary changing operators appear as unitary operators that map whole sectors of the extended Hilbert space with particular boundary conditions to sectors with different boundary conditions.\cite{affleck1994} Details of this analytic continuation can also be found in Appendix \ref{Appendix:KeldyshEntangle} and in Refs.[\onlinecite{calabrese2005,calabrese2006}].

Folding the system on the $(x,\tau)$-world sheet, the local quenching problem can be regarded as a boundary conformal field theory with a changing boundary condition, Neumann to Dirichlet at the origin. Such a boundary conformal field theory was studied in the context of the X-ray edge singularity and quantum wires.\cite{affleck1994,wong1994} The Neumann boundary condition implies that the odd boson is continuous at $x=0$ while the Dirichlet boundary condition implies that the odd boson is discontinuous at $x=0$. As discussed by Affleck and Ludwig,\cite{affleck1994} this can be written as:
\begin{eqnarray}
	\nonumber && \phi_{o}(0^{+},t< 0) = \phi_{o}(0^{-},t < 0) \\
	&& \\
	\nonumber && \phi_{o}(0^{+},t > 0 ) = - \phi_{o}(0^{-},t > 0 )
 \end{eqnarray}
This has an interpretation of a branch cut running on the $(x,\tau)$-world sheet across which the odd boson is discontinuous. In Ref.[\onlinecite{affleck1994}] it was noted that this branch cut corresponds to an insertion of a dimension $h=\frac{1}{16}$ operator at the quench times, $\tau_{1},\tau_{2}$. 

Observing that the boundary conditions are independent of the radius and following arguments by Ref.[\onlinecite{affleck1994}], it is suggestive that there is dimension $h=\frac{1}{16}$ twist operator in the odd boson theory (independent of the boson radius) which changes the boundary condition from Neumann to Dirichlet. Given such generality, one wonders if there is a larger algebraic structure describing the local quench which includes the boundary condition changing operator in its operator algebra. In the sequel, we show that this structure is described by an orbifold conformal field theory.

\section{Extension to other Laughlin states}
\label{section:Orbifolds}

Much of the previous discussion was limited to the $\nu=1/2$ case and was special in that it could be formulated in terms of Majorana fermions. We argued previously that to generalize the previous results to the other Laughlin states, one needs to understand the theory in terms of free chiral bosons and to work with an extended Hilbert space where a dimension $h=\frac{1}{16}$ boundary condition changing operator is included. Such a Hilbert space is described by an orbifold conformal field theory, but before delving into orbifold theories, we need to describe the initial Hilbert space of the system which we will extend in the next section.

The total charge of the system is conserved in the quenching process so that physical charged excited states are those preserving the total charge. Initially, one knows that in the limit of two decoupled edge states, the charged excitations that conserve total charge are processes where an electron tunnels from one edge to the other, $\mathcal{L}_{int} = \Psi^{\dagger}_{L}\Psi_{R} + \textrm{h.c.}$ In the even and odd basis, one can identify the vertex operator $e^{\pm i\sqrt{2g}\phi_{o}}$ with $g=m$ as generating these excited states. By charge conservation, the even sector does not have charged excited states, and its Hilbert space is generated by the $U(1)$ current. However, charged excited states are allowed in the odd sector by electron tunneling, so the physical Hilbert space is generated by the $U(1)$ current as well as the vertex operator $e^{\pm i \sqrt{2m}\phi_{o} }$.

Representations of this Hilbert space can be constructed with the requirement that they be local with respect to the generators $j=\partial\phi_{o}$ and $V_{\pm} = e^{\pm i \sqrt{2m}\phi_{o} }$. This construction is well known and extends the $U(1)$ Kac-Moody (KM) algebra to conformal field theories described by the $A_{m}$-series. The primaries in the theory can be found in any standard text.\cite{bigyellowbk,ginsparg1990}
\begin{eqnarray}
 \textrm{\bf 1}  \quad &\leftrightarrow& \quad   h =0  \nonumber \\\nonumber \\
 		j \quad &\leftrightarrow& \quad  h=1  \nonumber \\ \nonumber \\
		\phi_{m} \quad &\leftrightarrow& \quad h= \frac{m}{4} \nonumber  \\ \nonumber \\
		\phi_{k} \quad &\leftrightarrow &\quad h = \frac{k^{2} }{4m}, \,\,\,\,\, k = 1, \,.\,.\,. m-1 
\end{eqnarray}
where $\phi_{m} = e^{\pm i \sqrt{2m}\phi}$ and $\phi_{k} = e^{\pm i k/\sqrt{2m} \phi_{o} }$.

\subsection{Orbifolds}
The twist operator $\sigma$ exists explicitly in the fermion theory as spin operators, but by considering a $\mathbb{Z}_{2}$ orbifold of the free Gaussian theory, a twist operator can be included in the boson theory. On the $(x,\tau)$-world sheet of the odd boson, one has that the $\phi_{o} (z)= -\phi_{o}(z)$ as $z$ approaches the branch cut running between the two quench times. Imposing $\phi_{o}(0^{+})=-\phi_{o}(0^{-} )$ on the $(x,\tau)$-sheet has consequences for the target space.\cite{dixon1987} For a compact boson, $\phi_{0} \in S^{1}=[0,2\pi R]$, this means that the target space is identified under the action of the discrete group $\mathbb{Z}_{2}$. The theory must include an operator which takes $\phi_{o}\rightarrow -\phi_{o}$.

As argued before, the theory describing the odd boson is a conformal field theory in the $A_{m}$-series. To this theory, we wish to allow for an anti-periodic boson and need to identify the target space under the group $\mathbb{Z}_{2}$. One needs to consider the orbifold $A_{m}/\mathbb{Z}_{2}$. The partition function for orbifold theories under the group $\mathbb{Z}_{N}$ was found by Ginsparg.\cite{ginsparg1988} Decomposing this partition function into $A_{m}$ characters and observing the transformation properties under the modular $\mathcal{S}: \tau \rightarrow -1/\tau$ and modular $\mathcal{T}: \tau \rightarrow \tau+1$ transformation allows one to deduce the primaries, the modular $S$-matrix of the theory and fusion rules.\cite{dijkgraaf1989,bigyellowbk} The primaries of the theory are,
\begin{eqnarray}
\nonumber	\textrm{\bf 1} \quad &\leftrightarrow& \quad h = 0 \\ \nonumber \\
 \nonumber j  \quad &\leftrightarrow& \quad  h=1  \\ \nonumber \\
\nonumber	\phi_{m}^{(i)} \quad &\leftrightarrow& \quad  h= \frac{m}{4}, \,\,\,\,\,\, i = 1,2 \\ \nonumber  \\
\nonumber	\phi_{k}   \quad &\leftrightarrow &\quad  h = \frac{k^{2} }{4m}, \,\,\,\,\, k =1, \,.\,.\,. N-1 \\ \nonumber \\
\nonumber	\sigma_{1,2}  \quad &\leftrightarrow& \quad  h = \frac{1}{16}  \\ \nonumber \\
	\tau_{1,2}  \quad &\leftrightarrow& \quad  h = \frac{9}{16} 
\end{eqnarray}
The $\mathbb{Z}_{2}$ transformation $g$ acts on these representations as $g: [\phi_{k}] \rightarrow [\phi_{2m-k}]$ and so there are two fixed points $[\phi_{0}]$ and $[\phi_{m}]$. The operators $\phi_{m}^{(i)}$ correspond to a sector invariant under $\phi_{o}\rightarrow -\phi_{o}$ (i.e. $\phi_{m}^{(1)}=\cos (\sqrt{2m} \phi_{o})$ )and a sector that is broken by the transformation $\phi_{o}\rightarrow -\phi_{o}$ (i.e. $\phi_{m}^{(2)} = \sin(\sqrt{2m}\phi_{o})$ ). Physically, one can understand them as the tunneling and current operators  for the odd boson respectively. The operator $\phi_{k}$ are invariant under the transformation $\phi_{k} = \cos(\frac{k}{\sqrt{2m} } \phi_{o} )$. \cite{dijkgraaf1989} $\sigma_{1,2}$ correspond to the trivial and non-trivial representations of $\mathbb{Z}_{2}$. For $m$ even, it turns out that the $\sigma_{i}$ fields are self conjugate while for $m$ odd, $\sigma_{1}$ and $\sigma_{2}$ are each others conjugates.\cite{dijkgraaf1989} Using the fusion rules (see Appendix \ref{appendix:OrbifoldAlgebra}) and decomposing the partition function $Z_{DD}$ in terms of the characters  of the $A_{m}/\mathbb{Z}_{2}$ theory, it can be seen that the $\sigma_{1,2}$ operators act as boundary changing operators. For $m$ even, fusing with $\sigma_{1}$ gives $Z_{DD} \rightarrow_{\sigma_{1} }  \, Z_{ND} \rightarrow_{\sigma_{1}} \, Z_{NN}$ while for $m$ odd one has $Z_{DD} \rightarrow_{\sigma_{1} } \, Z_{ND} \rightarrow_{\sigma_{2}} \, Z_{NN}$.\cite{affleck2001}

For $\nu=1/2$, the orbifold theory is $A_{2}/\mathbb{Z}_{2}$ which is isomorphic to the tensor product of two $c=\frac{1}{2}$ Virasoro algebras ({\it i.e.\/} two decoupled Ising models).\cite{dijkgraaf1989} This was exactly the behavior observed before. The partition function for the odd boson one observed could be decomposed into a product of $c=\frac{1}{2}$ partition functions. In addition to this, the primaries of the theory are
\begin{eqnarray}
\nonumber	\textrm{\bf 1} \quad & \leftrightarrow& \quad  h = 0 \\ \nonumber \\
 \nonumber 		j \quad &\leftrightarrow &\quad  h=1  \\ \nonumber \\
\nonumber		\phi^{(i)}_{2} \quad& \leftrightarrow& \quad  h= \frac{1}{2}  \\ \nonumber \\
\nonumber		\phi_{1} \quad &\leftrightarrow& \quad  h = \frac{1 }{8} \\ \nonumber \\ 
\nonumber		\sigma_{1,2} \quad&\leftrightarrow& \quad h = \frac{1}{16}  \\ \nonumber \\
		\tau_{1,2} \quad &\leftrightarrow &\quad  h = \frac{9}{16} 
\end{eqnarray}
We see that the algebra naturally includes the twist fields $\sigma_{1,2}$ and $\tau_{1,2}$ and that there is an operator 
corresponding to the quasiparticle tunneling operator,
\begin{equation}
	\begin{array}{ccccc}
	j \times j = 1&  \phi_{2}^{(i)} \times \phi_{2}^{(i)} = 1 & \phi_{2}^{(1)} \times \phi_{2}^{(2)} = j \\ \\
	 \, & \sigma_{i}\times\sigma_{i} = 1 + \phi_{2}^{(i)}  & \\ \\
	 \,&  \sigma_{1}\times \sigma_{2} = \phi_{1} &
	 \end{array}
\end{equation}
The operator $\phi_{1}$ has a fusion rule consistent with its interpretation as $\cos( \frac{k}{2} \phi_{o})$.\cite{dijkgraaf1989}

In addition, we see that the dimension $h=\frac{1}{8}$ field we can regard as the dimension of the two spin field composite. Hence, one of the $\sigma$'s originally at the center can be taken to be in the trivial representation $\sigma_{1}$ of $\mathbb{Z}_2$ non-local with respect to the fermion field $\eta_{1}$ and the other in the non-trivial representation and non-local with respect to the fermion $\eta_{2}$ (see Fig. \ref{fig:boundary-changing}). When the interaction strength is switched, precisely one spin field interacts with either $\eta_{1,2}$ giving a twisted boundary condition. This behavior was also seen in the odd boson partition function and explicit calculation in Section \ref{section:Majorana}.

For $\nu=1/3$, the theory is described by the orbifold theory $A_{3}/\mathbb{Z}_2$ which corresponds to the $\mathbb{Z}_4$ parafermion theory.\cite{dijkgraaf1989} This can also be seen by comparing the fusion algebra and operator content of the two theories:
\begin{eqnarray}
\nonumber \\ 
\nonumber		\textrm{\bf 1}   \quad &\leftrightarrow& \quad h = 0 \\ \nonumber\\
\nonumber 		j  \quad &\leftrightarrow &\quad h=1 \\ \nonumber \\
\nonumber		\phi^{(i)}_{3} \quad &\leftrightarrow & \quad  h= \frac{3}{4} \\ \nonumber \\
\nonumber		\phi_{1} \quad &\leftrightarrow& \quad h = \frac{1 }{12} \\ \nonumber \\ 
\nonumber		\phi_{2}  \quad &\leftrightarrow& \quad  h = \frac{1 }{3} \\ \nonumber \\
\nonumber		\sigma_{1,2}  \quad &\leftrightarrow& \quad  h = \frac{1}{16} \\ \nonumber \\
		\tau_{1,2}  \quad &\leftrightarrow& \quad  h = \frac{9}{16} 
\end{eqnarray}
The fusion rules here are somewhat different than for $m$ even. The operator algebra of \textrm{\bf 1}, $j$ and $\phi_{3}^{(i)}$ now has a $\mathbb{Z}_4$ symmetry, and it is given by
\begin{equation}
	\begin{array}{ccccc}
	j \times j = 1&  \phi_{3}^{(i)} \times \phi_{3}^{(i)} = j & \phi_{3}^{(1)} \times \phi_{3}^{(2)} = 1 \\ \\
	 \, & \sigma_{i}\times\sigma_{i} =  \phi_{3}^{(i)} + \phi_{1}  & \\ \\
	 \,&  \sigma_{1}\times \sigma_{2} =1+ \phi_{2} &
	 \end{array}
\end{equation}
The fusion algebra for the $\phi_{k}$ vertex operators is unchanged.\cite{dijkgraaf1989} The fusion rules and modular $S$-matrices for other filling fractions are given in general in Appendix \ref{appendix:OrbifoldAlgebra}.

\subsection{Calculation of the Noise}

\label{section:Noise}

One can then go about computing the noise in the backscattering current in the language of bosons. In the original left-right basis, the backscattered current is given by
\begin{equation}
	I_{b} = (\rho_{L}^{in} + \rho_{R}^{in}) - (\rho_{L}^{out} + \rho_{R}^{out})
\end{equation}
 $\rho_{i}^{out}$ are charge densities measured at a position after the point contact while $\rho_{i}^{in}$ are charged densities measured at a position before the point contact. In the rotated even-odd basis, the backscattering current can be written as, $I_{b} = \frac{\sqrt{2\nu }}{2\pi} (\partial_{x}\phi_{-}^{o} - \partial_{x}\phi_{+}^{o})$ where $\phi_{\pm}$ denotes after and before the point contact at $x=0$. To connect with the boundary conformal field theory formalism, one can fold the system and regard $\partial \phi (x<0,t)$ as a right moving wave $\bar{\partial}\bar{\phi}(x>0,t)$. The backscattering current is then $I_{b} =\sqrt{\frac{\nu}{8\pi^{2}} }(\bar{j}^{o} - j^{o})$ where $x>0$ now. This form has the advantage of explicitly involving a primary field of the theory. Intuitively, the backscattered charge is simply the difference between the (incoming) left moving charge density and the (outgoing) right moving charge density. 
The noise is,
\begin{widetext}
\begin{eqnarray}
	\exptvalue{ \{ I_{b}(\tau_{2}), I_{b}(\tau_{1}) \} }_{a}&& = \exptvalue{j^{o}(i\tau_{1}+\delta_{\epsilon}) j^{o}(i\tau_{2}+\delta_{\epsilon})}_{a} + \exptvalue{\bar{j}^{o}(i\tau_{1}-\delta_{\epsilon} ) \bar{j}^{o}(i\tau_{2}-\delta_{\epsilon} )}_{a} \nonumber \\
	&&- \exptvalue{j^{o}(i\tau_{1}+\delta_{\epsilon} ) \bar{j}^{o}(i\tau_{2}-\delta_{\epsilon} )}_{a} - \exptvalue{\bar{j}^{o}(i\tau_{1}-\delta_{\epsilon}) j^{o}(i\tau_{2}+\delta_{\epsilon})}_{a} + (\tau_{1}\leftrightarrow \tau_{2})
\end{eqnarray}
\end{widetext}
where $\exptvalue{...}_{a}$ denotes a correlation function with boundary condition $a$ at $x=0$ (Neumann/fixed or Dirichlet/free) and fixed (Neumann) at $x=\ell$. Here $\tau_{1} = t_{1}-i\epsilon$ and $\tau_{2}=t_{2}+i\epsilon$ when $t_{1}$ is on the top contour and $t_{2}$ on the bottom contour and vice versa in the second contribution. $\delta_{\epsilon}$ is a short distance cutoff, $\delta_{\epsilon} \rightarrow 0$. In the presence of a boundary condition $a$ these correlation functions have been computed using methods of conformal invariance and their relationship to the modular $S$-matrix of the bulk theory were found in Refs.[\onlinecite{affleck1993, cardy1991}],
\begin{equation}
	\exptvalue{ \phi (z_{1} ) \phi(z_{2} ) }_{a} = \exptvalue{ \phi (z_{1} ) \phi (z_{2} )} \Bigg\{ \begin{array}{ll} 1&\textrm{for}\; xy >0 \\  \frac{S_{a}^{\Delta}/ S_{0}^{\Delta} }{S_{a}^{0}/S_{0}^{0} } &\textrm{for}\; xy<0 \label{eqn:BoundaryBulkCorr}
	\end{array}
\end{equation}
where $z_{1} = i \tau_{1} + x$ and $z_{2} = i\tau_{2} + y$ and $\phi$ is a primary field with scaling dimension $\Delta$. By knowing which primary field corresponds to boundary condition $a$ one can use Cardy's formula to compute the correlation function in the presence of the boundary. 

For $\nu=1/2$, we showed that the theory can be thought of as $A_{2}/\mathbb{Z}_2$ which is isomorphic to a theory of two decoupled Ising models. The boundary state corresponding to free boundary conditions (Dirichlet) is associated with the primary $\sigma$ while the fixed boundary conditions (Neumann) is associated with the identity or $\epsilon$ primary operators. Using the $S$-matrix for the Ising model, its easy to see that the correlation functions with boundary condition $a$ have the prefactors, 
\begin{eqnarray}
\nonumber		\textrm{fixed/N: } \quad &\Rightarrow& \quad  \frac{S_{0}^{1}/ S_{0}^{1} }{S_{0}^{0}/S_{0}^{0} } =1 \\ \nonumber \\
		\textrm{free/D: } \quad &\Rightarrow &\quad \frac{S_{1/16}^{1}/ S_{0}^{1} }{S_{1/16}^{0}/S_{0}^{0} }  = -1 
\end{eqnarray}
For other filling fractions, the identification of boundary states and boundary conditions is more complicated but we find that a similar structure is present.

Since the characters  for the  $A_{m}/\mathbb{Z}_2$ theory are known, the partition functions $Z_{NN}$ and $Z_{ND}$ can be decomposed into a sum of characters from which one may identify the boundary states.\cite{oshikawa2001,affleck2001}  Note that here we worked with the T-dual ($r \rightarrow \frac{2}{r}$) of the theory  in Refs.[\onlinecite{oshikawa2001,affleck2001}] so that $D\leftrightarrow N$. For $m$ even, Neumann and Dirichlet boundary conditions, it was found that the orbifold boundary states are,
\begin{widetext}

\begin{eqnarray}
	\nonumber \state{N } = \frac{1}{^{4}\sqrt{2 m} } \left( \state{0} + \state{j} +\frac{1}{\sqrt{2} } \left( \state{ \phi_{m}^{(1)} }+ \state{ \phi_{m}^{(2)} } \right) + \sqrt{2} \sum_{k=1}^{m-1} \state{\phi_{k} } \right)  \\ 
	&&  \\
	 \nonumber \state{D} =  \frac{1}{^{4}\sqrt{2 m} } \left( \sqrt{m} \state{0} - \sqrt{m} \state{m} + \sqrt{\frac{m}{2} } \left( \state{\phi_{m}^{(1)}} - \state{\phi_{m}^{(2)} } \right) \right) 
\end{eqnarray}

\end{widetext}

For $m$ odd, the analogous states can be written down using the appropriate $S$-matrix (see Appendix \ref{appendix:OrbifoldAlgebra}). From the modular $S$-matrix, one then can identify the Neumann boundary condition with the highest weight state $\phi_{k}$ and the Dirichlet (free) boundary condition with the $\sigma$ field. Note that the boundary states above are {\it not} the usual boundary states but are $\mathbb{Z}_2$ invariant linear combinations.\cite{affleck2001} 
However, the fact remains that they transform in the $[\phi_{k}]$ and $[\sigma]$ representations respectively. For $\nu=1/m$ one finds that it is generally true that
\begin{eqnarray}
\nonumber		\textrm{fixed/N: } \quad &\Rightarrow& \quad \frac{S_{k}^{1}/ S_{0}^{1} }{S_{k}^{0}/S_{0}^{0} } =1 \\ \nonumber \\
		\textrm{free/D: } \quad &\Rightarrow &\quad  \frac{S_{1/16}^{1}/ S_{0}^{1} }{S_{1/16}^{0}/S_{0}^{0} }  = -1 
\end{eqnarray}
For arbitrary $\nu$, one finds that in the limit $\epsilon\rightarrow 0, \delta_{\epsilon} \rightarrow0$, the noise from the local quench is given by
\begin{equation}
	\exptvalue{ \{ I_{b}(\tau_{2}), I_{b}(\tau_{1}) \} }  = \frac{\nu}{2\pi^{2} }  \frac{1}{(\delta + i (t_{1}-t_{2}) )^{2} }
\end{equation}
in agreement with the calculation for $\nu=1/2$ so that the second cumulant has a similar form as (\ref{eqn:2ndcumulant}).

A non-trivial check of these results can also be done by considering the boundary condition on the odd currents. In terms of the odd currents, the reflecting and transmitting boundary condition were given by $j^{o} = \bar{j}^{o}$ and $j^{o} = -\bar{j}^{o}$ respectively.\cite{wong1994} 
Imposing either of these boundary conditions, one sees that the left-left and right-right correlation functions are unaffected but the left-right correlation function changes by a minus sign.
\begin{eqnarray}
	j^{o}=\bar{j}^{o} ; && \exptvalue{j^{o}(z_{1}) \bar{j}^{o}(\bar{z}_{2}) } = \frac{1}{2} \frac{1}{(z_{1}-\bar{z}_{2} )^{2}} \nonumber \\ 
	&& \\
	j^{o}=-\bar{j}^{o}  ; && \exptvalue{j^{o}(z_{1}) \bar{j}^{o}(\bar{z}_{2}) } = -\frac{1}{2}\frac{1}{(z_{1}-\bar{z}_{2} )^{2}} \nonumber
\end{eqnarray}
so that when the point contact is closed, the noise measured is zero while when it is opened, the noise is non-zero. This is in agreement with the results using the operator algebra of the $A_{m}/\mathbb{Z}_2$ orbifold. Note that this correlation function is special in that the boundary condition involves the primary fields $j^{o}$ and $\bar{j}^{o}$ that appear in the desired correlation function but Eqn.(\ref{eqn:BoundaryBulkCorr}) is more general than this.

While the second cumulant can easily be shown to grow logarithmically in time, our result shows that the coefficient of this logarithm {\em depends on the modular $S$-matrix elements of the theory and not just to its central charge}, so that even for $c=1$ the second cumulant does {\it not} fully give the entanglement entropy. 

\subsection{Calculation of the full counting statistics}
\label{section:FCS}

The full counting statistics (FCS) is the generating function for the moments of the charge transfered between two reservoirs through a constriction. In Ref.[\onlinecite{KlichLevitov}], the FCS was calculated for the process of opening and closing the QPC, with the result that the distribution function for the transmitted charge is purely Gaussian. Here we shall extend our methods developed in the previous sections to find the FCS in the interacting case, and conclude that the non-interacting result is in fact robust to the presence of interactions. It is plausible that even in the interacting case, there exists a simple relation between the FCS and the entanglement entropy.

To avoid boundary effects, it is convenient for our purposes to consider a Corbino (annular) geometry, with a single QPC inducing backscattering between the edges. In this new geometry, we should actually reverse the protocol appearing in previous sections: first the QPC is open, then is quickly closed, and remains in this state during a time period $\Delta t$, at the end of which it is opened again. Going to the odd/even basis for the two edge bosons, the charge carried by the odd boson, $Q_o=\int dx \rho_o=\frac{\sqrt{\nu}}{2\pi}(\phi_o(0^+)-\phi_o(0^-))$, is controlled by the singularity induced by the branch cut for the closed QPC (assumed to be located at $x=0$). The full counting statistics is then given by the following correlation function,
\begin{eqnarray}
	\chi(\lambda)=\left\langle\left\{e^{i \lambda Q_o(\Delta t)},e^{-i\lambda Q_o(0)}\right\}\right\rangle
\end{eqnarray}
which, using the observation above, can be reduced to the calculation of vertex operators at the vicinity of the QPC. This leads to the main result,
\begin{eqnarray}
	\chi(\lambda)=\exp\left\{-\frac{\lambda^2}{2} \frac{\nu}{2\pi^2}\left[\log\left(\frac{\delta^2+\Delta t^2}{\delta^2}\right)\right]\right\}
\end{eqnarray}
which is the typical generating function for a Gaussian distribution. Consequently, only the second cumulant ({\em i.e.\/} the noise) is non-zero. Interestingly, except for the explicit dependence on the filling factor $\nu$, the full counting statistics has the same form as the non-interacting electron case. 

Writing the boson in the even and odd basis and folding the system, this problem can be mapped onto the boundary sine-Gordon model. In the basis of kink and anti-kink solitons of charge $q$ and $-q$ respectively, distribution functions can be found via the thermodynamic Bethe ansatz for the entire range of $\lambda$.\cite{FendleyLudwigSaleur-noise,KomnikSaleur-full} The problem is then similar to the non-interacting electron problem only the particles are collective modes rather than the original electrons. However, in this picture, the full counting statistics is difficult to compute. The reflection and transmission coefficients of the time dependent scattering matrix\cite{levitov1993,KlichLevitov} are difficult to write in the kink and anti-kink basis. 

\section{Energy-Momentum noise in the Quantum Ising model}

\label{section:Thermal}
Although in previous sections we showed that the noise generated by a quench can also evolve logarithmically with time, we will show now that this is not always true. This is in contrast to the static entanglement entropy which depends logarithmically on the size of the observed region\cite{Callan1994,Vidal2003,Calabrese2004} and the dynamic entanglement entropy which evolves logarithmically with time.\cite{calabrese-2009b}

Consider a one-dimensional quantum Ising model that can be split in two by changing with time the strength of just one link.
This effectively implies a change in boundary conditions at that link. We use the Majorana fermion description of the critical quantum Ising model, 
\begin{eqnarray}
	\mathcal{L}=i \tilde{\eta}_R(\partial_t-\partial_x)\tilde{\eta}_R+i\eta_L(\partial_t-\partial_x)\eta_L
\end{eqnarray}
where $\eta_L$ ($\eta_R$) is a left-going (right-going) Majorana fermion, and $\tilde{\eta}_R(x)=\eta_R(-x)$ was flipped. At the center of the Ising chain ($x=0$) we modify the link strength, which translates in the effective Fermion description to a coupling to the local energy density operator $\epsilon=\eta_L\eta_R$ at the same point 
\begin{eqnarray}
	\mathcal{L}'=i\lambda\delta(x)\eta_L\eta_R
\end{eqnarray}
This marginal perturbation leads to the following relation between the fields on the two sides of the impurity
\begin{eqnarray}
	\left(\begin{array}{c} \tilde{\eta}_R(0^+)\\ \eta_L(0^+)\end{array}\right)=M \left(\begin{array}{c} \tilde{\eta}_R(0^-)\\ \eta_L(0^-)\end{array}\right)
\end{eqnarray}
with
\begin{eqnarray}
	M=\frac{1}{1+\left(\frac{\lambda}{2}\right)^2}\left(\begin{array}{cc} 1-\left(\frac{\lambda}{2}\right)^2 & -\lambda \\ \lambda & 1-\left(\frac{\lambda}{2}\right)^2\end{array}\right)
\end{eqnarray}
When $\lambda$ is fine tuned to the point $\lambda=2$, $M$ becomes completely off-diagonal $M\to -i\sigma_y$. When $\lambda\to 0$, $M$ approaches the identity matrix, $M\to I$.

The quantum Ising model and its effective field theory, the free Majorana fermion, has a (discrete) global $\mathbb{Z}_2$ symmetry. Contrary to the Luttinger-type models which have a globally conserved $U(1)$ charge and a locally conserved (dimension one) charge current, in the case of the quantum Ising model the only global conservation law is the energy-momentum and its locally conserved (dimension two) energy-momentum current. In contrast with the Luttinger model where the quantum quench does not change the total charge, in the Ising model the quantum quench changes the Hamiltonian and hence the total energy. Although the energy-momentum tensor of the Majorana fermion field theory (and of the Ising model) remains locally conserved everywhere {\em except} at the location of the quantum impurity, these physical differences lead to a distinct asymptotic time dependence of the noise in the energy-momentum current shown below.

The energy density along the edge is defined by
\begin{eqnarray}
	\rho_E(x)=\eta_L i\partial_x \eta_L-\tilde{\eta}_Ri\partial_x\tilde{\eta}_R
\end{eqnarray}
The rate of back-scattering of energy by the modified link is
\begin{eqnarray}
	I_E=\rho_E(0^+)-\rho_E(0^-)
\end{eqnarray}
Using the relations above, for the case that the link is fully transmitting ($\lambda=0$), the heat current is zero as the contributions on the two sides of the link cancel. In the limit of a fully reflecting link, we get $\rho_E(0^+)=-\rho_E(0^-)$, and the two contributions add up.

Since the scaling dimension of the energy current is $2$ we get the following expression for the thermal noise $E_2$:
\begin{eqnarray}
	\nonumber & E_2=\frac{1}{\pi^2}\int_0^{\Delta t} dt_1 dt_2 \left(\frac{1}{t_1-t_2+i \delta}\right)^4\\
	&=\dfrac{1}{\pi^2}\dfrac{3\delta^2 \Delta t^2+\Delta t^4}{3\delta^2(\delta^2+\Delta t^2)^2}
\end{eqnarray}
For large $\Delta t$, the thermal noise approaches a non-universal, cutoff dependent, value
\begin{eqnarray}
	E_2\sim \frac{1}{3 \pi^2\delta^2}+\frac{1}{\pi^2}\frac{1}{\Delta t^2}
\end{eqnarray}

\section{Conclusions}

In the previous sections we identified an extended theory of the bosonic Luttinger description of the quantum Hall edge states in the presence of an impurity. The extended theory, being an $A_m/\mathbb{Z}_2$ orbifold theory (for $\nu=1/m$), explicitly contains an operator inducing the transition between the zero-backscattering and fully-backscattering fixed points of the impurity. In particular, for $\nu=1/2$ the extended theory decouples into two copies of $\mathbb{Z}_2$ parafermions, whereas for $\nu=1/3$ it coincides with $\mathbb{Z}_4$ parafermions instead. The presence of the primary field in the algebra which induces the transition between the two fixed points at the impurity demonstrates that in the sudden approximation equilibrium is maintained through each process of opening or closing of the QPC, and that no transient effects are expected. 

Using these methods we calculated the FCS for the interacting Luttinger liquid for the process of a sudden opening (at $t_0$) then closing (at $t_1$) of the QPC. Only the second cumulant of the transferred charge is nonzero, and its logarithmic dependence on $t_1-t_0$ suggests that a relation between the entanglement entropy and the FCS should exist even in the interacting case. In that sense, the noise ``measures'' the entanglement entropy. The only difference from the non-interacting case is the explicit appearance of the filling factor in the noise.

In a more general setting, however, the relation between the FCS and the entanglement entropy does not seems to be structural.  First, it requires the presence of a conserved current with dimension 1 (essential for a logarithmic behavior of the correlation function), and of an associated and strictly conserved global charge. In contrast,  the energy-momentum noise through a weak link in the one-dimensional Ising model does not reproduce, or even has the same form,  as the entanglement entropy. We discuss this case briefly in Section \ref{section:Thermal}.  Also, for any quantum Hall state which contains neutral edge modes, the generated entanglement entropy is proportional to the total central charge, $c$, but in the absence of an impurity the noise involves only the pure current present in the charge carrying $c=1$ theory alone. This can be amended by adding the central charge explicitly into the relation between the noise and the entanglement entropy, but this argument seems to be ignoring the mechanisms that create entanglement in the neutral theory. Moreover, even for the simple Luttinger case the central charge does not appear in the noise (which depends only on the fusion rules and the conformal dimensions of the CFT) while it certainly appears explicitly in the entanglement entropy. Indeed, our results show that the noise generated is a measurement of \emph{dynamical} entanglement and not of the \emph{static} entanglement that is associated with the observation of a subsystem in an eigenstate of the full system. \cite{calabrese-2009b}

Interestingly, we also find that the time dependent correlation functions have a universal scaling behavior as in the case of the global quench.\cite{calabrese2006}  However, the universal behavior is characteristic of an orbifold theory of the original system. It may be of interest to study this experimentally where corrections to scaling are sure to occur. The flow away from criticality would presumably involve the more exotic fusion rules. 

In summary, in this paper we have computed the noise of the tunneling current generated by a quantum quench of a point contact in a fractional quantum Hall fluid. This was done by explicit solution for the bosonic Laughlin state at $\nu=1/2$ and by conformal field theory methods for a general Laughlin state. We have also computed the growth of the entanglement entropy due to the quantum quench. We found that even though the time dependence of the noise has the {\em same form} as the entanglement entropy the latter has a more intricate dependence on the properties of the conformal field theory and the Hilbert space of the edge state. This result suggests that the Klich-Levitov protocol may not generally supply a procedure to measure the entanglement entropy. The question of measuring the entanglement entropy remains open.

\begin{acknowledgments}
We thank Ian Affleck, John Cardy, Claudio Chamon, Paul Fendley, Israel Klich, Leonid Levitov, and Chetan Nayak  for enlightening discussions and suggestions.  This work was supported in part by the
National Science Foundation grant DMR 0758462 at the University of Illinois (EF and BH).  EG was supported by the Institute for Condensed Matter Theory of the University of Illinois.
\end{acknowledgments}

\appendix

\section{Details of $\nu=1/2$}
\label{appendix:Nu-half}

In this appendix we give some details of the re-fermionization technique used in the exact solution of the $\nu=1/2$ (bosonic) quantum Hall effect. For $\nu=1/2$, the Lagrangian for the odd boson in the presence of a QPC at $x=0$ is given by ($v\equiv 1$)
\begin{eqnarray}
	\mathcal{L}=\frac{1}{4\pi}\partial_x\phi_o(\partial_t-\partial_x)\phi_o+\lambda\delta(x)e^{i\phi_o(x)}+\mathrm{h.c.}
\end{eqnarray}
where the odd density $\rho_o(t,x)=[\rho_R(t,x)-\rho_L(t,-x)]/\sqrt{2}$. We refermionize by defining \cite{kane1992}
\begin{eqnarray}
	\psi(t,x)=\frac{1}{\sqrt{2\pi}}:e^{i \phi_o(t,x)}:
\end{eqnarray}
so that $\{\psi(t,x),\psi^\dag(t,x')\}=\delta(x-x')$ holds. The full Hamiltonian can now be written as
\begin{eqnarray}
	\nonumber H_o&=&\int dx \psi^\dag(x)i\partial_x\psi(x)\\
	&&\quad\quad+\sqrt{2\pi}\delta(x)\left[\lambda \psi(x)\gamma+\lambda^* \gamma\psi^\dag(x)\right]
\end{eqnarray}
Note that $\rho_o(x)=\psi^\dag(x)\psi(x)$, and $\{\gamma,\gamma\}=2$. Here $\gamma$ is a Klein factor which appears in a careful handling of the zero modes of the boson field.

Assuming for simplicity $\lambda=\lambda^*\in \mathbb{R}$, the equations of motion can be found and are given by
\begin{eqnarray}
	\nonumber &&i\partial_t\psi=i\partial_x\psi-\sqrt{2\pi}\lambda\gamma\delta(x)\\
	&&i\partial_t\psi^\dag=i\partial_x\psi^\dag+\sqrt{2\pi}\lambda\gamma\delta(x)\\
	\nonumber &&i\partial_t\gamma=2\sqrt{2\pi}\lambda(\psi^\dag(0)-\psi(0))
\end{eqnarray}
The equations of motion can be solved by expanding the fields in modes as (see e.g. \onlinecite{sandler-1999})
\begin{eqnarray}
	\nonumber &&\psi(x,t)=\sum_\omega e^{i\omega(x+t)}\left\{\begin{array}{cc} A_\omega & x<0 \\ B_\omega & x>0 \end{array}\right.\\
	&&\psi^\dag(x,t)=\sum_\omega e^{i\omega(x+t)}\left\{\begin{array}{cc} A^\dag_{-\omega} & x<0 \\ B^\dag_{-\omega} & x>0 \end{array}\right.\\
	\nonumber &&\gamma(t)=\sum_\omega e^{i\omega t}\gamma_\omega
\end{eqnarray}
The modes satisfy the following relations, derived by integrating the equations of motions around $x=0$,
\begin{eqnarray}
	\nonumber &i(B_\omega-A_\omega)=\sqrt{2\pi}\lambda\gamma_\omega\\
	&i(B_{-\omega}^\dag-A_{-\omega}^\dag)=-\sqrt{2\pi}\lambda\gamma_\omega\\
	\nonumber &\!\!\!\!\omega\gamma_\omega=2\sqrt{2\pi}\lambda\left[\frac{1}{2}(A_\omega+B_\omega)-\frac{1}{2}(A_{-\omega}^\dag+B_{-\omega}^\dag)\right]
\end{eqnarray}
Defining $\omega_B\equiv 4\pi\lambda^2$, we get
\begin{eqnarray}
	\nonumber \left(\begin{array}{c}B_\omega \\ B^\dag_{-\omega}\end{array}\right)&=&\frac{1}{i\omega-\omega_B}\left(\begin{array}{cc} i\omega & -\omega_B \\ -\omega_B & i\omega\end{array}\right)\left(\begin{array}{c}A_\omega \\ A^\dag_{-\omega}\end{array}\right)\\
	&\equiv& M_\omega\left(\begin{array}{c}A_\omega \\ A^\dag_{-\omega}\end{array}\right)
\end{eqnarray}
So that when $\omega_B\to 0$ (fully transmitting QPC), $M_\omega\to I$, while when $\omega_B\to\infty$ (fully reflecting QPC) $M_\omega\to\sigma_x$.
\\

\section{Modular $S$-matrix and Fusion Rules at Level $1/\nu=m$ }
\label{appendix:OrbifoldAlgebra}
With the convention that the matrix takes the vector labelled by $(1,j,\phi_{m}^{(j)}, \phi_{k'},\sigma_{j},\tau_{j})$ to the same vector with $j\rightarrow i$, $k\rightarrow k'$ and $\sigma_{ij}=2\delta_{ij}-1$, the modular $S$-matrix for $m$ even can be written as:
\begin{widetext}
\begin{equation}
	S_{m,\textrm{even} } = \frac{1}{\sqrt{8m} } \left( \begin{array}{cccccc}
	1 & 1 & 1 & 2 &\sqrt{m} & \sqrt{m} \\
	1 & 1 & 1 & 2 &- \sqrt{m} & -\sqrt{m} \\
	1 & 1 & 1 & 2 (-1)^{k'} & \sigma_{ij}\sqrt{m} & \sigma_{ij} \sqrt{m}\\
	2 & 2 & 2(-1)^{k} & 4 \cos \pi\frac{kk'}{2m} & 0 & 0 \\
	\sqrt{m} & -\sqrt{m} &\sigma_{ij} \sqrt{m} & 0 & \delta_{ij}\sqrt{2m} &-\delta_{ij} \sqrt{2m} \\
	\sqrt{m} & -\sqrt{m} &\sigma_{ij} \sqrt{m} & 0 & -\delta_{ij}\sqrt{2m} &\delta_{ij} \sqrt{2m} \\
	 \end{array} \right)
\end{equation}
while for $m$ odd, the modular $S$-matrix takes a slightly different form.
\begin{equation}
	S_{m,\textrm{odd} } = \frac{1}{\sqrt{8m} } \left( \begin{array}{cccccc}
	1 & 1 & 1 & 2 &\sqrt{m} & \sqrt{m} \\
	1 & 1 & 1 & 2 &- \sqrt{m} & -\sqrt{m} \\
	1 & 1 & -1 & 2 (-1)^{k'} & i\sigma_{ij}\sqrt{m} & i\sigma_{ij} \sqrt{m}\\
	2 & 2 & 2(-1)^{k} & 4 \cos 2\pi\frac{k k'}{2m} & 0 & 0 \\
	\sqrt{m} & -\sqrt{m} &i\sigma_{ij} \sqrt{m} & 0 & e^{i \pi\sigma_{ij}/4} \sqrt{2m} &- e^{i \pi\sigma_{ij}/4} \sqrt{2m} \\
	\sqrt{m} & -\sqrt{m} &i\sigma_{ij} \sqrt{m} & 0 & - e^{i \pi\sigma_{ij}/4}\sqrt{2m} & e^{i \pi\sigma_{ij}/4}\sqrt{2m} \\
	\end{array}
	 \right)
\end{equation}
\end{widetext}

For $m$ even, the elements $\{1, j, \phi_{m}^{(i)}\}$ form a $\mathbb{Z}_2 \times \mathbb{Z}_2$ subalgebra. One has
\begin{eqnarray}
	&& j \times j = 1 \nonumber \\
	&& \phi_{m}^{(i)} \times  \phi_{m}^{(i)} = 1 \\
	&& \phi_{m}^{(1)} \times \phi_{m}^{(2)} = j \nonumber
\end{eqnarray}
The twist operators have the fusion rules
\begin{eqnarray}
	&& \sigma_{i} \times \sigma_{i} = 1 + \phi_{m}^{(i)} + \sum_{k,even} \phi_{k} \nonumber \\
	&&  \sigma_{1}\times\sigma_{2} = \sum_{k,odd} \phi_{k}
\end{eqnarray}
For $m$ odd, the elements $\{1, j,\phi_{m}^{(i)}\}$ form a $\mathbb{Z}_4$ subalgebra. 
\begin{eqnarray}
	&& \nonumber j \times j =1 \\
	&& \phi_{m}^{(1)} \times \phi_{m}^{(2)} = 1 \\
	&& \nonumber \phi_{m}^{(i)} \times \phi_{m}^{(i)} = j
\end{eqnarray}
and for the twist fields,
\begin{eqnarray}
	&& \nonumber \sigma_{i}\times \sigma_{i} = \phi_{m}^{(i)} + \sum_{k,odd} \phi_{k} \\
	&& \sigma_{1}\times \sigma_{2} = 1 + \sum_{k,even} \phi_{k}
\end{eqnarray}
The fusion rules for the vertex operators $\phi_{k}$ are the same for $m$ even or odd.
\begin{eqnarray}
	&& \nonumber \phi_{k} \times \phi_{k'} = \phi_{k+k} + \phi_{k-k'} \\
	&&  \phi_{k} \times \phi_{k} = 1 + j + \phi_{2k} \\
	&& \nonumber \phi_{m-k} \times \phi_{k} = \phi_{2k} + \phi_{m}^{(1)} + \phi_{m}^{(2)} \\
	&& \nonumber j\times \phi_{k} = \phi_{k}
\end{eqnarray}
The fusion rules for the $\tau_{i}$ fields can be found easily by applying $j\times \sigma_{i} = \tau_{i} $.

\section{Schwinger-Keldysh and Entanglement Entropy}
\label{Appendix:KeldyshEntangle}

The non-equilibrium problem of locally quenching the system at the point contact is best dealt with in terms of the density matrix formalism. Initially, the system has a density matrix $\rho_{0}=\state{\psi_{0}(x)}\astate{\psi_{0}(x)}$ where the states $\state{\psi_{0}(x)}$ are eigenstates of the Hamiltonian with Neumann boundary conditions imposed at the point contact and ends of the system. At time $t=0$, the point contact is opened and the boundary condition at the boundary is changed to Dirichlet.

After the quench, the density matrix is given by
\begin{widetext}
\begin{equation}
	\langle{\phi'(x)}\vert \rho(t)\vert \phi(x) \rangle = \astate{\phi'(x)} U^{\dagger}(t) \rho_{0} U(t) \state{\phi(x)} = \langle{\phi'(x)}\vert e^{-it H - \epsilon H } \vert\psi_{0}(x) \rangle \langle \psi_{0}(x)\vert e^{it H - \epsilon H} \vert \phi(x) \rangle
\end{equation}
\end{widetext}
where the regulator $\epsilon$ has been included to adiabatically cut-off the high momentum modes. Here, $H$ is the Hamiltonian with transmitting boundary condition. The first term describes forward evolution from the initial state to the final state $\state{\phi(x)}$ while the second describes the time reversed evolution back to the initial state.

Following Refs.[\onlinecite{calabrese2005,calabrese2006}], the Schwinger-Keldysh time contour can be thought of as the time contour for a conformal field theory. On the forward branch, one continues $\tau = -\epsilon- it$, $\phi(x,t) \rightarrow \phi(x,\tau)$. Now, $\tau \rightarrow -i\tau$, Wick rotating, this can be thought of as propagating the initial state for time $\tau$ to the final state.
\begin{equation}
	 \langle{\phi'(x)}\vert e^{-it H - \epsilon H } \vert\psi_{0}(x) \rangle =  \langle{\phi'(x)}\vert e^{ -i H \tau} \vert \psi_{0}(x) \rangle 
\end{equation}
One can think of the density matrix in terms of a path integral, but to do so, a complete set of states needs to be introduced. Here we work with the {\it extended} Hilbert space which is spanned by states with all possible boundary conditions. Inserting a complete sets of states $\int d\phi_{i} \,\, \vert \phi_{i} \rangle \langle \phi_{i}\vert$ at each time slice $\delta \tau_{i}$ and also a complete set of conjugate states $\int d\Pi_{i} \,\, \vert \pi_{i} \rangle \langle \pi_{i}\vert$.
\begin{widetext}
\begin{eqnarray}
	\nonumber  \langle{\phi(x)}\vert e^{ -i H \tau} \vert\psi_{0}(x) \rangle && = \int \prod_{i=1}^{N} d\phi_{i} \prod_{i=1}^{N} d\pi_{i} \,\, e^{i \delta \tau_{i} H(\pi_{i},\bar{\phi_{i}} ) } \, \, e^{i \pi_{i+1} \phi_{i+1} } \,\, e^{-i\pi_{i+1} \phi_{i} } \\
	&& = \int D\phi \,\, e^{  \int d\tau dx \,\, \mathcal{L}[\phi(x,\tau) ] } \,\, \delta(\phi(x,\tau=0) - \phi(x))\, \, \delta( \phi(x,\tau_{1}) - \psi_{0}(x) )
\end{eqnarray}
\end{widetext}
Now, $d\tau = -i dt$ so that this is the Euclidean path integral with the boundary condition that at $\tau_{1} = -\epsilon -it$, the field takes value $\phi(x,\tau_{1}) = \psi_{0}(x)$ and $\phi(x,\tau)=\phi(x)$ at $\tau=0$. A similar set of manipulations can be performed for the backward branch as well. At the point $t=0$, they differ on a set of measure zero so that the path integral can be thought of as over the entire surface parameterized by $(x,\tau)$ with changing boundary conditions at $\tau_{1}$ and $\tau_{2}$. At the end of the calculation, one should continue back to real time $t$ and take the limit $\epsilon \rightarrow 0$.\cite{calabrese2005} Here, before the quench $t<0$, we apply reflecting boundary conditions and after the quench, we have transmitting boundary conditions. In the sequel, we denote this surface, with changing boundary conditions in time at $x=0$ as $\Sigma$.

To compute the entanglement entropy, one uses the replica trick to compute, $\textrm{Tr }\rho_{A}^{n}$. The reduced density matrix can be visualized as stitching together the cylinder on the $B$ side while leaving the $A$ side open. Multiple copies of $\rho_{A}$ are stitched cyclically on the cut. \cite{calabrese2005, calabrese2006} In Ref.[\onlinecite{Calabrese2004}], it was shown using general transformation properties of the stress energy tensor that this branch cut amounts to a twist operator $\Phi_{n}(z)$ insertion on the Riemann surface with conformal dimension, $\Delta_{n} = \frac{c}{24}(n-1/n)$. The stitched surface amounts to computing the correlation function of the twist operator.
\begin{equation}
	\textrm{Tr }\rho_{A}^{n} = \exptvalue{\Phi_{n}(z=0) }_{\Sigma}
\end{equation}
Here $\exptvalue{ ... }_{\Sigma}$ denotes an average taken over the surface with changing boundary conditions. To compute this correlation function, we use the Zhukowski mapping which avoids the algebraic structure discussed in the text. The advantage is that including an $n$-twist field as part of an extended Hilbert space is difficult and is left as an open question here. 
Using the mapping,
\begin{equation}
	w=\frac{1}{\epsilon}\left(z+\sqrt{z^2+\epsilon^2}\right)
\end{equation}
the surface $\Sigma$ is mapped to the right half plane with a single boundary condition, $\Sigma'$. The correlation function on $\Sigma$ is related to the correlation function on $\Sigma'$ 
\begin{equation}
	\exptvalue{\Phi_{n}(z) }_{\Sigma} = \abs{\frac{dw}{dz} }^{-\Delta_{n} } \exptvalue{\Phi_{n}(w) }_{\Sigma'}
\end{equation}
Differentiating with respect to $n$ at $n=1$ yields the entanglement entropy
\begin{equation}
	S(t) = \frac{c}{6} \log \left( \frac{t^{2} +\epsilon^{2} }{a \epsilon/2} \right)+ \tilde{c}_{1}
\end{equation}
where $\tilde{c}_{1}$ is a non-universal constant and $a$ a lattice cutoff.\cite{calabrese2005} In the limit $t\gg \epsilon$ the entanglement entropy becomes
\begin{equation}
	S (t) = \frac{c}{3} \log\frac{t}{a} + \tilde{k}_{1} 
\end{equation}
so that the amount of entanglement grows logarithmically in time after the quench at $t=0$ and is proportional to the central charge.


\end{document}